# Spontaneous rotation of a nanosatellite FITSAT-1

Y. Kawamura[1] and T. Tanaka[2]

[1]Department of Intelligent Mechanical Engineering, Fukuoka Institute of Technology, 3-30-1 Wajirohigashi, Higashiku, Fukuoka, 811-0295, Japan

[2]Department of Information Engineering, Fukuoka Institute of Technology, 3-30-1, Wajirohigashi, Higashiku, Fukuoka, 811-0295, Japan

## Abstract

Spontaneous rotation of an ultra-small satellite was observed and its driving torque was explained by the thermal interaction between the air molecules and the surfaces of the satellite heated by the radiation from the earth. This mechanism has the similarity with a usual "radiometer", except the point that the velocity of the satellite is sufficiently faster than that of the thermal velocity of the air molecules, and that the mean free path of the air molecules is sufficiently longer than the characteristic length of the satellite. Using dimension analysis, the torque was found to be significant to a small size of satellite. This rotation mechanism can be applied to any small objects, which are revolting around a planet radiating a black body radiation and has the atmospheric gas.

## 1. Introduction

We developed a small cube satellite [1] named FITSAT-1 [2-5] and released it from the ISS (International Space Station) on October the 5th in 2012. We observed a small amount of gradual increase in the angular velocity of the satellite. The driving torque, which induce the automatic rotation of the satellite we observed, was explained using a modified radiometer model ("space radiometer" model). It can be applied to a moving object in the space, whose velocity is sufficiently faster than the thermal velocity of the ambient gas molecules and the characteristic length of the body is shorter than the mean free path of the gas. This rotation mechanism can be applied to any small objects, which are revolting around a planet radiating a black body radiation and has the atmospheric gas around it, such as space debris in case of the earth.

## 2. Spontaneous rotation of a nanosatellite

Figure 1 shows the small satellite used in this experiment. (code name: FITSAT-1, nickname: "NIWAKA"), which was released from the ISS (International Space Station) on October 5th, 2012. It is a small cubic type satellite with the dimensions of 10cm × 10cm × 10cm and the weight of 1.3 kg. The moment of inertia around $z$ axis is $2.2 \times 10^{-3}$ kgm². It has no automatic control system, but has only a permanent magnet in it to direct the $z$ axis to the direction the earth magnet field, like a magnetic compass. Therefore, the satellite can rotate around the axis of the permanent magnetic ($z$ axis), if the torque around is applied.

After the launching of the satellite from the ISS, it took several tens of days to stabilize the initial movements around three freedoms of rotational motions, which were generated by the launcher of the ISS. These initial rotations were dumped by the eddy current induced in the aluminum panels of the satellite body induced by the interaction with the earth magnet. Dumping effect of the eddy current is not effective only to the rotation around the $z$ axis, because the $z$ axis is aligned to the direction of the earth magnet and the magnetic flux density of the earth magnet does not change by the rotation around the $z$ axis.

After the initial rotational motion of the satellite was stabilized, we noticed gradual acceleration of the rotation speed around the $z$ axis of the satellite, and it was accelerated gradually to be about 0.1 Hz for about three months. The rotational direction of the satellite was measured to be opposite to the revolution direction around the earth. The satellite was not equipped with the gyroscope. These measurement (rotational speed and direction) was performed by analyzing the periodical changes in the signal intensity of the power generation voltage of four pairs of solar cells, which are attached on four side panels of the satellite perpendicular to the $z$ axis.

These data could not be transmitted in real-time, but they could be obtained as the recorded data, when the satellite passes above the ground station at Fukuoka. The rotation velocity was also measured by the periodical changes in the signal intensity of the microwave from the satellite.

Figure 2 shows the increase in the angular velocity of the satellite for about three months, from the start of the measurement (November the 2nd, 2012) to February the 2nd, 2013. The plots fit to a dashed line is the data from the November 18th, 2012 to February 10th , 2013.

## 3. Rotation mechanism

After the consideration of several models, we made a physical model to explain this acceleration phenomenon of the rotational speed quantitatively, as shown in Fig. 2. Here, the rotation torque is generated by the momentum exchange between the air molecules and the satellite.

In this model, it is the first necessary conditions that the velocity of the object ($U$) is sufficiently faster than the thermal velocity of the air molecules ($v_1$, $v_2$), and the mean free path length ($\lambda$) is sufficiently longer than the characteristic length of the object ($a$). The first necessary condition is written by

$$U \gg v_1, v_2 \quad \text{and} \quad a \ll \lambda \tag{1}.$$

Under this condition, the rear surfaces (C, D) are not hit by the air molecules, but the front surfaces (A, B) are hit by them and are effective to the generation of the torque.

The second necessary condition is that there must be temperature difference between the upper surface and lower surface of the object. It is a reasonable assumption, because the lower two surfaces (B, C) are heated by the radiation from the earth (300K), while the upper two surfaces (A, D) are cooled by the radiation to the space (3K). $T_1$ and $T_2$ are the surface temperatures of the upper panels (A, D) and lower panels (B, C), respectively. The second necessary condition is written by

$$\Delta T = T_2 - T_1 > 0 \qquad (2),$$

where $\Delta T$ is the temperature difference between these two.

$U$ is calculated to be about to be $7.2 \times 10^3$ m/s. $v_1$ and $v_2$ are the translational thermal velocities of the air molecules, and determined by the equations,

$$v_1 = \sqrt{\frac{3kT_1}{m}} , \quad v_2 = \sqrt{\frac{3kT_2}{m}} \qquad (3)$$

where, $k$ is the Boltzmann constant ($1.38 \times 10^{-23}$ J/K) and $m$ is the mass of the air molecules ($4.78 \times 10^{-26}$ kg). Considering $\Delta T \ll T_1$, $T_2$, the difference between $v_1$ and $v_2$ is approximated to be,

$$v_1 - v_2 \cong \frac{3}{2} \times \frac{\Delta T}{T_1} v_1 \qquad (4).$$

The direction of the rebounding molecules has the statistical angular distribution in practice, but in this analysis, in order to simplify the calculation, the rebounding of molecules is assumed to be a specular reflection as shown in Fig. 3. $F_1$ and $F_2$ are the reaction forces induced by the repulsion of the molecules. They are parallel to each other and have opposite directions. Considering the momentum exchange between the molecules on the surface of the satellite in a unite second, these are calculated to be

$$F_1 = \rho U S v_1 \cos\theta , \; F_2 = \rho U S v_2 \sin\theta \qquad (5),$$

where $\rho$ is the density of the air and estimated to be $4.4 \times 10^{-12}$ kg/m$^3$ using the table from U.S. Standard Atmosphere [6], and $S$ is the surface area of each panels ($S = a^2 = 0.01$m$^2$). Using the equations (3), torque around the $z$-axis is calculated to be

$$\tau = \frac{a}{2} \cdot F_1 \sin\theta - \frac{a}{2} \cdot F_2 \cos\theta = \frac{\rho U a^3}{4}(v_1 - v_2)\sin 2\theta \qquad (6).$$

Substituting the equation (4) to the equation (6), torque around the $z$-axis is expressed by

$$\tau = \frac{3\rho U v_1 a^3 \Delta T}{8 T_1}\sin 2\theta \qquad (7).$$

The system has a rotational symmetry for every 90 degrees, therefore the clockwise torque is continuously applied to the satellite. Integrating $\tau$ from $\theta = 0$ to $\theta = 90$ degrees, averaged torque is calculated to be

$$\overline{\tau} = \frac{3\rho U v_1 a^3 \Delta T}{4\pi T_1} \qquad (8).$$

The equation of motion for the rotation is given by,

$$\overline{\tau} = J\dot{\omega} \qquad (9),$$

where $J$ is the momentum inertia of the satellite around the $z$ axis, and is $2.2 \times 10^{-3}$ [kgm$^2$]. Using the equation (8) and (9), the relation between the angular acceleration $\dot{\omega}$ and the temperature difference $\Delta T$ can be expressed by

$$\dot{\omega} = \frac{3\,\rho\;Uv_1a^3}{4\,\pi\,T_1J}\;\Delta T \tag{10}$$

In this experiment the temporally averaged value of the temperature of the satellite was measure to be about 270 K, therefore $T_1$ was estimated to be 270 K considering $T_1 \cong T_2$. The sampling time of the temperature measurement of the panel was as slow as 5 minutes, therefore we could not identify $\Delta T$ experimentally [4]. Therefore, we assumed the temperature difference between two surfaces is calculated to be $\Delta T = 5$ K. (We think that $\Delta T = 5$ K is reasonable value, judging from the fact that the difference between the maximum and minimum temperature orbiting around the earth was measured to be about 25 K [4].) Substituting $T_1$=270K, $\rho = 4.4 \times 10^{-12}$ kg/m$^3$, $U = 7.2 \times 10^3$ m/s, $v_1 \cong 4.8 \times 10^2$ m/s, $J$ =2.2 $\times 10^{-3}$ kgm$^2$, $a =$ 0.1 m and $\Delta T = 5$ K into the equation (10), the angular acceleration of the satellite is calculated to be $\dot{\omega} = 3.1 \times 10^{-8}$ rad/s$^2$, which is shown using the dashed line and shows good agreement with the experimental results in Fig.2.

## 4. Discussion

The characteristic length of the satellite has the dimension of $[L^1]$ and the momentum inertia of the satellite have the dimension of $[L^5]$, where $L$ is the dimension of length. The density of the air $\rho$, the velocity of the satellite $U$, thermal velocity of the molecules $v$ and the temperature $T$ is independent from the dimension of the satellite, and they are considered to be constant value in this analysis. Therefore, the angular acceleration is calculated to be inversely proportional to the square of the characteristic length $[L]$, and written by

$$\dot{\omega} \propto L^{-2} \qquad\qquad (11)$$

This size effect will be the reason, why we found the automatic rotation phenomenon for the first time in our ultra-small satellite. Another reason may be that our satellite did not have any automatic control system for the rotational motion, but only has the permanent magnet to operate as the magnetic compass to keep the $z$ axis roughly perpendicular to the moving direction of the satellite.

In order to explain the automatic acceleration phenomenon, another several models were considered other than the "space radiometer model" mentioned above.

The first trial was the explanation using the radial density gradient of the air, which induces pressure difference between the upper surfaces (A in Fig. 3) and the lower surfaces (B in Fig. 3) and generates the torque around $z$ axis. In this model, the rotation direction becomes opposite to the experimental results, and the calculated value of the rotational acceleration torque was smaller than the measured value by three orders of magnitude.

The second trial was the explanation using the thermal expansion of the body of the satellite, which generates the difference in the forces applied on the upper surface (A) and lower surface (B). In this model, the direction of the rotation coincides with the experimental result, but the value of the rotational acceleration becomes smaller than the measured values by two orders of magnitude.

The third trial was the attribution to the photon pressure [7]. In this case, the axial asymmetry of the photon absorption coefficient on the surface is essential, which our satellite did not have.

The fourth trial was the explanation using the electromagnetic force generated by the interaction between the earth's magnet and the current generated by the solar cells. In this case, the rotation around $z$ axis, which coincides with the direction of the external magnetic field, cannot be induced according to the fundamental electromagnetic theory.

We could not find any reasonable explanation except the "space radiometer model" described here, and only this model could explain the phenomenon qualitatively and quantitatively.

## 5. Conclusions

Automatic rotation of an ultra-small satellite was observed. The driving torque induce the rotation was explained by the thermal interaction between the air molecules and the surfaces of the satellite heated by the radiation from the earth. Using dimension analysis, the torque was found to inversely proportional to the square of the dimension of the satellite and significant to a small size of the satellite. This rotation mechanism can be applied to small objects, which are revolting around any planet radiating a black body radiation and has the atmospheric gas.

## References
[1]. Heidt, H. and Puig-Suari, J., Moore, S. A., Nakasuka, S. & Twiggs, J. R. A new generation of picosatellite for education and industry low-cost space experimentation. 14th Annual/USU Conference on Small Satellites (SSC00-V-5) (2000).

[2]. Tanaka, K., Tanaka, T. & Kawamura, Y. Development of the Cubesat FITSAT-1. Proceedings of UN-Japan Nano-Satellite Symposium, NSS-04-0209, Nagoya (Japan), Oct.10, 11 (2012).

[3]. Mizoguchi, Y., Feng, K., Soda, T., Otsuka, T., Kinoshita, T., Nishimoto, K., Kawamura, Y., & Tanaka, T.  Observation of the LED signal from FITSAT-1. UN-Japan Nano-Satellite Symposium, NSS-04-0111, Nagoya (Japan), Oct.10, 11 (2012).

[4]. Tanaka, T., Kawamura, Y. & Tanaka, K. Development and Operations of Nano-Satellite FITSAT-1 (NIWAKA).  Acta Astronautica, Vol. 107, 112-129 (2015).

[5]. Kawamura, Y. & Tanaka, T. Transmission of the LED light from the space to the ground. AIP Advances, Vol.3, 102110 (2013), DOI:10.1063/1.4824853.

[6]. U.S. Standard Atmosphere, 1976. US government, Printing office, (1976).

[7]. Kershner, B. R. Satellite Rotation by Radiation Pressure. US Patent 3,145,948 (1964).

[8]. Cho, M., Masui, H., Akahoshi, Y., Hiraki, K., Iwata,M., Toyoda, K. & Hatta, S. Initial Operations of Center for Nanosatellite Testing. Proceedings of the 2nd Nano-Satellite Symposium (2011).

## Acknowledgements

Authors thank JAXA (Japan Aerospace Exploration Agency) for the continuing supports to perform "FITSAT-1" project, and also thank Center for Nanosatellite Testing, Kyushu Institute of Technology, for the thermal vacuum test and the vibration test of FITSAT-1 [8].

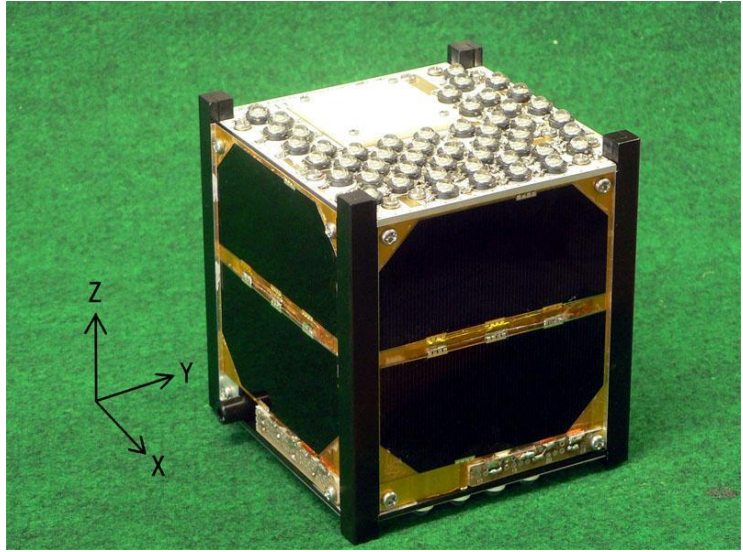

**Fig. 1** A small cube satellite (code name: FITSAT-1, nickname: "NIWAKA"). 8 solar panels of the same size are installed axial symmetrically on $X$(+) surface, $X$(-), surface $Y$(+) surface and $Y$(-) surface.

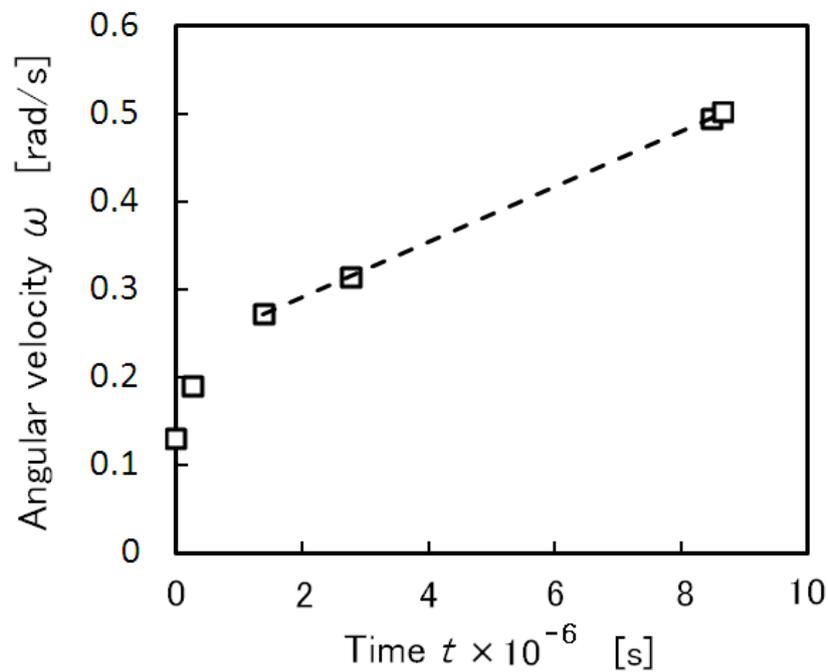

**Fig. 2** Gradual increase in the angular velocity of the satellite. The dotted line in the figure is the best-fit line of the data for 85 days (Nov. 2, 2012~Feb. 10, 2013), and has the slope of $3.1 \times 10^{-8}$ rad/s$^2$.

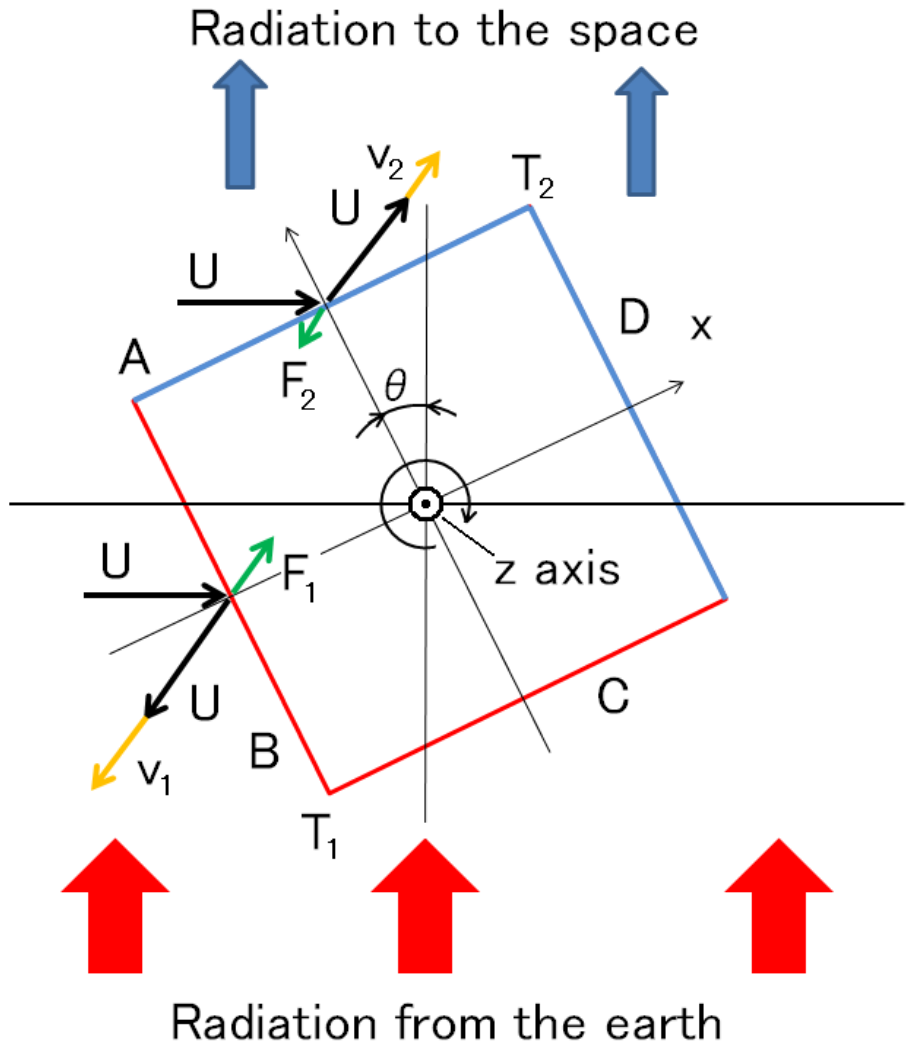

**Fig. 3** Automatic rotation model of the satellite by the momentum exchange between the air molecules and the heated surfaces.